\documentclass[aps,prd,preprint,tightening]{revtex4}
\usepackage{epsf,epsfig,graphics,graphicx,amsmath}
\usepackage[dvipsnames,usenames]{color}
\usepackage{comment}
\usepackage{ amssymb }

\newcommand{\vx}{\ensuremath{\vec{x}}}

\newcommand{\be}{\begin{equation}}
\newcommand{\ee}{\end{equation}}
\newcommand{\bea}{\begin{eqnarray}}
\newcommand{\eea}{\end{eqnarray}}
\begin{document}

\title{Power spectrum of primordial perturbations during ultra-slow-roll inflation with back reaction effects}

\author{Shu-Lin Cheng$^1$}
\email{shlcheng@gate.sinica.edu.tw}
\author{Da-Shin Lee$^2$}
\email{dslee@gms.ndhu.edu.tw}
\author{Kin-Wang Ng$^{1,3}$}
\email{nkw@phys.sinica.edu.tw}
\affiliation{$^1$Institute of Physics, Academia Sinica, Taipei 11529, Taiwan, R.O.C.}
\affiliation{$^2$Department of Physics, National Dong Hwa University, Hualien 97401, Taiwan, R.O.C.}
\affiliation{$^3$Institute of Astronomy and Astrophysics,
Academia Sinica, Taipei 11529, Taiwan, R.O.C. }

\date{\today}

\begin{abstract}
We develop a nonperturbative method through the Hartree factorization to   examine the  quantum fluctuation effects on the single-field inflationary models in a spatially flat FRW cosmological space-time. Apart from the background field equation as well as the Friedmann equation with the corrections of quantum field fluctuations, the modified Mukhanov-Sasaki equations for the mode functions of the quantum scalar field are also derived by introducing the nonzero $\Delta_B$ term. We consider the Universe undergoing the slow roll (SR)-ultra slow roll (USR) -slow roll (SR) inflation where
in particular the presence of the USR inflation triggers the huge growth of $\Delta_B$ that in turn gives the boost effects to the curvature perturbations for the modes that leave horizon in the early times of the inflation.
However, the cosmic friction term in the mode equation given by the Hubble parameter presumably prohibits the boost effects. Here we propose two representative models to illustrate these two competing terms.
\end{abstract}

\maketitle

\section{Introduction}
Inflation is the leading paradigm that prevails in the evolution of the early universe.
 During  the stage of primordial
acceleration,  vacuum  fluctuations of the gravitational and matter fields
 are amplified to the cosmological scales, giving the approximately scale invariant spectrum of small density fluctuations on large scales, which are ultimately responsible for the CMB anisotropies and the large scale structure of our Universe \cite{KT}.

The recent detection of one of the gravitational wave events is speculated  due to the merging of two 30 $M_{\odot}$  black holes, which might be primordial black holes (PBHs), resulting in LIGO coalescences \cite{ABB}.
In addition, PBHs could comprise a considerable fraction of
the dark matter, and thus leave imprints throughout the history of the Universe \cite{CAR,MES,CAR1} (see Ref. \cite{SAS} for a review). There has been increasing interest in the possibility of breaking scale invariance on small scales with a spike in the power spectrum, resulting in structures such as PBHs. The models include trapped inflation \cite{CHENG},
Higgs inflation \cite{EZQcrit}, double inflation \cite{KAN}, radiative plateau inflation \cite{BAL} or some string realizations \cite{CLN,CIC,OZS,DAL}, to cite a few. To generate a spike in the power spectrum, the inflation scenario must deviate from the slow-roll (SR) regime that typically gives the scale invariant power spectrum.
Ultra-slow-roll (USR) inflation  has been proposed  as  a transient  phase of single-field inflation \cite{KIN,MAR,CHE,BYR}.
Nevertheless,  in this scenario
the very small inflaton velocity will be significantly influenced by  quantum kicks, giving its fluctuations comparable to the mean value. A natural framework to involve quantum kicks is through the stochastic inflation \cite{STA}. Several works \cite{FIR,PAT,BIA,EZQ} have been devoted to the study of this quantum noise effects and find a significant boost of PBH production.

Alternatively, the quantum fluctuation effects arising from the inflaton field itself to the curvature perturbations
through the quantum loop calculations are considered in Ref.~\cite{boyan1}.  It is known that in the case of minimally coupled massless inflaton fluctuations in de Sitter space-time they suffer from the infrared divergence~\cite{fordbunch}.
During the SR inflation where our Universe is in approximately de Sitter space-time,
the resulting scale invariant power spectrum introduces the strong infrared enhancement of the one-loop corrections in $\langle \varphi^2 \rangle $ given by the inflation field fluctuations $\varphi$ to be defined in the text later that in turn induce sizable effects to the slow roll parameters. Then the study is to extend to the USR inflation as a transition phase. In particular, when the Universe goes from the USR to SR regimes,
there exists the accompanying large value of $\langle \varphi^2 \rangle $
due to the combined effects of the strong infrared enhancement and the spike in the power spectrum \cite{weican}. In this work, we will adopt the approach to taking into account the quantum field fluctuations $\langle \varphi^2 \rangle $ self-consistently by implementing the Hartree factorization.
We then introduce two models
when with no inclusion of $\langle \varphi^2 \rangle $  they are consistent with the observations.
We plan to adopt the method developed here to see whether or not the quantum field fluctuations will give sizable effects in each of the models.

Our presentation is organized as follows. In next section, we introduce the single-field inflationary models in a metric of the perturbed spatially flat Friedmann-Robertson-Walker (FRW) cosmological space-time starting from
the
Arnowitt-Deser-Misner (ADM) form by following \cite{weican}.  We then separate  the classical homogeneous background field ($\Phi_0$)  from the
quantum field fluctuations ($\varphi$).
The background field  in the FRW metric with the quantum loop corrections through $\langle \varphi^2 \rangle$ is derived whereas
the scale factor follows the modified Friedmann equations also including the loop contributions.  The equations of motion for the mode functions of  $\varphi$ can be derived
from the quadratic terms of the field $\varphi$ in the Lagrangian density as in \cite{weican} where in this work we further implement the Hartree factorization for self-consistently taking into account $\langle \varphi^2 \rangle$.
The power spectrum of primordial perturbations in a spatially flat gauge
can be expressed in terms of the mode functions.
 In Sec.~\ref{sec3},  we introduce two models for the Universe undergoing the SR-USR-SR inflation to compute the power spectrum of the curvature perturbations, with which to further examine the effects of quantum field fluctuations to the power spectrum.
Concluding remarks and discussions are in Sec.~\ref{sec4}.

\section{Single-field inflation model}\label{sec2}
We consider the inflation model given by
\bea\label{action}
S=&& S_g+S_{\phi}\nonumber\\
=&&\frac{1}{2} \int d^4 x \sqrt{-g} R+\int d^4 x \sqrt{-g} \Bigg[ - \frac{1}{2} \;
\partial_{\mu} \phi \, \partial^{\mu}\phi  -V(\phi) \Bigg] \;
\eea
with  $M_{Pl}^{-2}= 8 \pi G_N$  set to be $M_{Pl}^{-2}=1$, which will be restored if necessary.
One shall start from the  metric in the Arnowitt-Deser-Misner (ADM) form in the $3+1$ decomposition with  ${\cal N}$ the lapse function and ${\cal N}^i$  the shift-vector as well as the 3 spatial dimensions metric $h_{ij}$~\cite{clu,kaz,weican}.
In particular, for a spatially flat
Friedmann-Robertson-Walker (FRW) cosmological space-time  ${\cal N}=1$, ${\cal N}_i=0$, and $h_{ij}=a^2 (t) \delta_{ij}$ where $a$ is a scale factor. In the ADM formalism, $h_{ij}$ and $\phi$ are dynamical variables whereas ${\cal N}$ and  ${\cal N}_i$ are Lagrangian multipliers due to the Hamiltonian and momentum constraints.
In the course of inflation, we consider the homogeneous mean field $\Phi_0$ and  the quantum fluctuations around it $\varphi(\vec{x},t)$ defined as
\be\label{tad}
\phi(\vec{x},t)= \Phi_0(t)+\varphi(\vec{x},t)\;,
\ee
\noindent with
\be\label{exp}
\langle
\varphi(\vx,t)\rangle =0 \;.
\ee
The expectation value is given by the non-equilibrium quantum state that later will
be specified to be the Bunch-Davies state often studied in the literature.
It is found more convenient to work in the spatially flat gauge by writing ${\cal N}=1+ \delta {\cal N}$
where $\delta  {\cal N}, {\cal N}_i$, and $\varphi$ are  perturbed fields. Notice that $\delta {\cal N}$ and ${\cal N}_i$ are all suppressed  not only by $1/M^2_{Pl}$ but also due to the smallness of $\dot\Phi_0$ during both SR or USR inflations, and they are
\be
\delta {\cal N} \propto \sqrt{\epsilon} \, , \quad \quad \quad {\cal N}_i \propto \sqrt{\epsilon} \, .
\ee
Henceforth, $\delta {\cal N}$ and ${\cal N}_i$ can be ignored  as compared with  loop effects from quantum  scalar field of the order of the power spectrum of the curvature perturbations $\propto H/(M_{Pl}\sqrt{\epsilon})$  where $H$ is the Hubble parameter and $\epsilon$ is the Hubble flow parameter being small during the USR and SR inflations \cite{weican}. {
The tadpole
method (see  Ref.~\cite{boyan3} and references
therein) then is implemented to derive the equation of motion with the loop corrections for the homogeneous
expectation value of the inflaton field from the action (\ref{action})  by requiring the condition $\langle
\varphi(\vx,t)\rangle =0 $:
\be \label{phi_loop}
\ddot{\Phi}_0+3\,H \,\dot{\Phi}_0+V'(\Phi_0)+\frac{1}{2} V'''(\Phi_0) \, \langle \varphi^2 (\vec x,t) \rangle =0 \, ,
\ee
where the dot means the time derivative of the cosmic time $t$.
The Hubble parameter $H$ obtained from the Hamiltonian constraint   with ${\cal N}=1$, ${\cal N}_i=0$ and $h_{ij}= a^2 \delta_{ij}$ giving the Friedmann equation becomes \cite{weican}
\bea \label{frieman_q}
H^2 &&=\frac{1}{3 M^2_{Pl}} \bigg\langle \frac{1}{2} {\dot{\phi}^2} +\frac{\partial_i \phi \partial_i \phi}{2 a^2} +V(\phi) \bigg\rangle \nonumber\\
&&=\frac{1}{3 M^2_{Pl}} \Big( \frac{1}{2} {{\dot\Phi_0}}^2  +V(\Phi_0)\Big) +\frac{1}{3 M^2_{Pl}} \bigg\langle \frac{1}{2} {\dot{\varphi}}^2 +\frac{\partial_i \varphi \partial_i \varphi}{2 a^2} +\frac{1}{2} V''(\Phi_0) \varphi^2+ \cdot \cdot\cdot \bigg\rangle \nonumber\\
&&= H_0^2 +\delta H^2 \, ,
\eea
where
\be \label{H0}
H_0^2=\frac{1}{3 M^2_{Pl}} \Big( \frac{1}{2} {\dot{\Phi_0}}^2  +V(\Phi_0)\Big)\,  ,
\ee
and the loop correction to the Hubble parameter  $\delta H^2$ is given by
\be \label{deltah}
\delta H^2=\frac{1}{3 M^2_{Pl}} \bigg\langle \frac{1}{2} {\dot{\varphi}}^2 +\frac{\partial_i \varphi \partial_i \varphi}{2 a^2} +\frac{1}{2} V''(\Phi_0) \varphi^2 \bigg\rangle \, .
\ee
{Additionally, the spatial Fourier transform of the field
operator $\varphi_k $ obeys the mode equations, which can be read off from the quadratic terms of $\varphi$ in the Lagrangian density (\ref{action}) with the further  Hartree factorization approximation, namely $\varphi^4\rightarrow 6 \langle \varphi^2 \rangle \varphi^2$,  as \cite{clu,kaz,weican}}
\begin{equation}
\ddot{\varphi_k} + 3H \dot{\varphi_k} + \left[\frac{k^2}{a^2}+ V''(\Phi_0) - \frac{1}{a^3 M_{P}^2}\frac{d}{dt}\left( \frac{a^3\dot{\Phi_0}^2}{H} \right) {+\frac{1}{2} V^{[4]}(\Phi_0) \, \langle \varphi^2 \rangle }\right]\varphi_k = 0 .
\label{phieom2}
\end{equation}
 However, $ \langle \varphi^2 \rangle$  has the usual ultraviolet divergence in the loop momentum integral that can be removed by a proper procedure of regularization/renormalization by defining the renormalized counterpart.  In the case of minimally coupled massless inflaton fluctuations in de Sitter space-time, it also suffers from the infrared divergence~\cite{fordbunch}. In Ref.~\cite{boyan1}, its infrared enhancement giving sizable effects to  the slow roll parameters is studied. Later the infrared enhancement is also explored in the USR inflation~\cite{weican}.
In this work, we self-consistently incorporate $\langle \varphi^2 \rangle$ into the dynamics where in particular the infrared momentum modes $(k \leq H)$ are taken into account by introducing the momentum cutoff $\Lambda=H$ in the momentum integration.

Next,  the Hubble flow parameters are defined as
\begin{equation}\label{Hubble_flow}
\epsilon = - \frac{\dot{H}}{H^2},\quad \eta = \frac{\dot\epsilon}{\epsilon H}.
\end{equation}
We can re-write Eq.~(\ref{phieom2}) in terms of $\zeta_k=\delta\varphi_k/(\sqrt{2\epsilon})$ as
\begin{equation}
\ddot{\zeta_k} +\left( 3 + \eta \right) H \dot{\zeta_k} + \left[ \frac{k^2}{a^2} + \Delta_B \right]  \zeta_k = 0,
\label{MS2a}
\end{equation}
where
\begin{eqnarray}
\Delta_{B} &=&\Delta_{B0} + { V^{[4]}(\Phi_0) \, \langle \zeta^2 \rangle \, \epsilon } \nonumber\\
\Delta_{B0} &=& V''[\Phi_0]-  \frac{1}{a^3 M_{P}^2}\frac{d}{dt}\left( \frac{a^3\dot{\Phi}_0^2}{H} \right) + \frac{H}{2}\dot{\eta} - \frac{H^2}{4}\eta^2 + \frac{3H^2}{2}\eta .
\label{Box1}
\end{eqnarray}
Notice that ignoring the quantum fluctuations corrections $\langle \varphi^2 \rangle$ or $\langle \zeta^2 \rangle$ of the redefined variable in $\Delta_B$ above, $\Delta_B \rightarrow \Delta_{B0}$. Then, using the equation of motion for $\Phi_0$ again by neglecting the contributions from quantum fluctuations in (\ref{phi_loop}) as well as the Friedmann equation for $H_0$  (\ref{H0}}) that
the dynamics of the background fields are taken into account only lead to $\Delta_{B0}=0$.  Then, the equation of motion for $\zeta$ in (\ref{MS2a}) reduces to the known  Mukhanov-Sasaki equation~\cite{msvar}.  Nevertheless, including the quantum fluctuations  will give the nonzero $\Delta_B$ in (\ref{Box1}) with the contributions  from the nonzero $\Delta_{B0}$ via (\ref{phi_loop}) and (\ref{frieman_q}), which implicitly have the quantum fluctuation corrections, and also from the term of $ V^{[4]}(\Phi_0) \, \langle \zeta^2 \rangle \, \epsilon $.
 Thus, $\Delta_{B} \neq 0$ becomes the indicator, giving the effects from the quantum fluctuations to the curvature perturbation to be discussed later.
\begin{comment}
 \textcolor{blue}{ Assuming to ignore $\langle \zeta^2 \rangle$, it is straightforward to see that Eq.~(\ref{MS2a})  reduces to the Mukhanov-Sasaki equation~\cite{msvar} the conformal time $\tau$.}\\\\
 \textcolor{red}{Shun-Lin:\\
 The equations for the cases without the backreaction effects  are:
\be
\ddot{\Phi}_0+3\,H_0 \,\dot{\Phi}_0+V'(\Phi_0)=0 \, ,
\ee
\be
H_0^2=\frac{1}{3 M^2_{Pl}} \Big( \frac{1}{2} {\dot{\Phi_0}}^2  +V(\Phi_0)\Big)\,  ,
\ee
\begin{equation}
\ddot{\zeta_k} +\left( 3 + \eta \right) H_0 \dot{\zeta_k} + \left[ \frac{k^2}{a^2} + \Delta_B \right]  \zeta_k = 0,
\label{MS2a}
\end{equation}
where
\begin{equation}
\Delta_B = V''[\Phi_0]-  \frac{1}{a^3 M_{P}^2}\frac{d}{dt}\left( \frac{a^3\dot{\Phi}_0^2}{H} \right) + \frac{H}{2}\dot{\eta} - \frac{H^2}{4}\eta^2 + \frac{3H^2}{2}\eta \ll 1 .
\end{equation}
Or
\begin{equation}
\ddot{\varphi_k} + 3H_0 \dot{\varphi_k} + \left[\frac{k^2}{a^2}+ V''(\Phi_0) - \frac{1}{a^3 M_{P}^2}\frac{d}{dt}\left( \frac{a^3\dot{\Phi_0}^2}{H_0} \right)\right]\varphi_k = 0 .
\label{phieom2}
\end{equation}
In particular, Eq.(15) can be rewritten as the Mukhanov-Sasaki equation given by
\begin{equation}
{\zeta^{''}_k} +\left( 2 + \eta \right) a H_0 {\zeta'_k} + k^2  \zeta_k = 0,
\end{equation}
in terms of the conformal time $\tau$.
They are the equations you solved in your previous paper to be compared with our current cases.}
\end{comment}
Here the initial conditions for $\zeta_k$ are chosen to be the standard Bunch-Davies vacuum states.  For the early times $|k\tau| \gg 1$ where $\tau$ is the conformal time, $d\tau = dt / a$,
the mode functions behave the same as free-field modes in the Minkowski
space-time, i.e.,
\begin{equation}
\zeta_k\to \frac{1}{\sqrt{2k}} e^{-ik\tau}\frac{1}{a\sqrt{2\epsilon}}.
\label{initialcon}
\end{equation}
As such the power spectrum of the curvature perturbation is given by
\begin{equation}
\Delta^2_{\zeta_k}=\frac{k^3}{2\pi^2} \left| \zeta_k \right|^2.
\label{deltazetak}
\end{equation}
When ignoring $\langle \varphi^2 \rangle$, for small $\epsilon$ and $\vert  \eta\vert $, the Hubble parameter
$H$  is approximately a constant where the scale factor is $a=e^{H_0t}=-1/(H_0\tau)$ in a era of de Sitter spacetime, and   Eq.~(\ref{MS2a}) has a simple analytic solution,
\begin{equation}
\zeta_k= \frac{\pi^{1/2}H_0}{2k^{3/2}}\frac{1}{\sqrt{2\epsilon}}(-k\tau)^{3/2} H^{(1)}_{3/2}(-k\tau),
\end{equation}
giving the power spectrum of the curvature perturbations
\begin{equation}
\Delta^2_{\zeta_k}=\frac{H_0^2}{8\pi^2 \epsilon}\, .
\label{deltaappro}
\end{equation}

In fact, in the theory of quantum fields in curved spacetime,
there is no unique choice of a vacuum state.
 In this article we focus on the standard
choice often adopted in the literature,  which allows us to include the
quantum corrections into the standard results in the literature. A study
of quantum loop corrections with different initial states is an important
aspect that deserves further study.
In the next section, we will study the effects of the back reaction of $\langle \varphi^2 \rangle $ to the curvature  perturbations.\\

\section{quantum fluctuations in SR-USR-SR inflation}\label{sec3}

In this section, we calculate the power spectrum of the curvature perturbation numerically.
We utilize the inflation models with an inflection point that will generate a large spike in the power spectrum as the source for the production of PBHs. Here we provide two models where if we ignore the effects of quantum fluctuations effects, the amplitude of the power spectrum of the curvature perturbation as well as the levels of the scalar spectral index, the tensor-to-scalar ratio, and the scalar spectral index running on large cosmological scales are all consistent with the Planck measurements:
$\Delta_{\zeta_k}^2\simeq 2.4\times 10^{-9}$, $n_s\simeq 0.97$, $r<0.06$, and $|dn_s/d\ln k|<0.013$~\cite{planckinflation,bk18}.
Using the Hartree factorization to self-consistently take into account  quantum field fluctuations, we  examine the backreaction effects of $\langle \varphi^2 \rangle$ to the dynamics.

 \subsection{Model 1}
\begin{figure}[htp]
\centering
\includegraphics[width=0.6\textwidth]{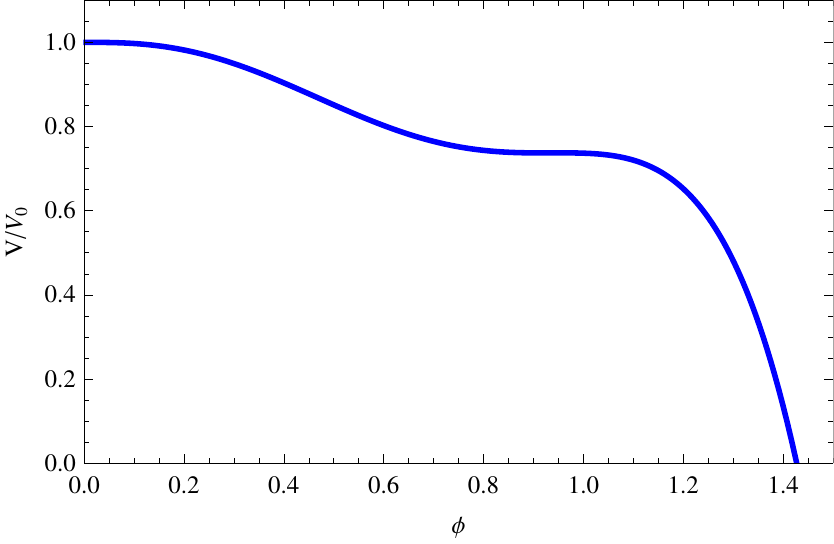}
\caption{The inflaton potential $V(\phi)$ in Eq.~(\ref{potential1}), normalized by $V_0$. All dynamical variables in this figure and in the following figures are rescaled by the reduced Planck mass, $M_p=2.435\times 10^{18}$ GeV. }
\label{figpotential}
\end{figure}

 The first model potential $V(\phi)$ is parametrized as
\begin{equation}
V(\phi) = V_0 \left( 1+c_1 \frac{\phi}{\Lambda} + \frac{c_2}{2}\frac{\phi^2}{\Lambda^2} + \frac{c_3}{3!}\frac{\phi^3}{\Lambda^3} + \frac{c_4}{4!}\frac{\phi^4}{\Lambda^4} + \frac{c_5}{5!}\frac{\phi^5}{\Lambda^5} \right) ,
\label{potential1}
\end{equation}
with
\begin{eqnarray}
&&c_\Lambda = 0.3,\quad V_0 = {1.0} \times 10^{-14}, \nonumber \\
&&c_1 = {-0.7276} \times 10^{-4},\; c_3 = -0.52,\; c_4 = 1.0,\; c_5 = -0.6407.
\end{eqnarray}
The typical profile of the $V(\varphi)$ is depicted in Fig.~\ref{figpotential}.
%Fig.1 potential

%Fig.2 phi and H
\begin{figure}[htp]
\centering
\includegraphics[width=0.6\textwidth]{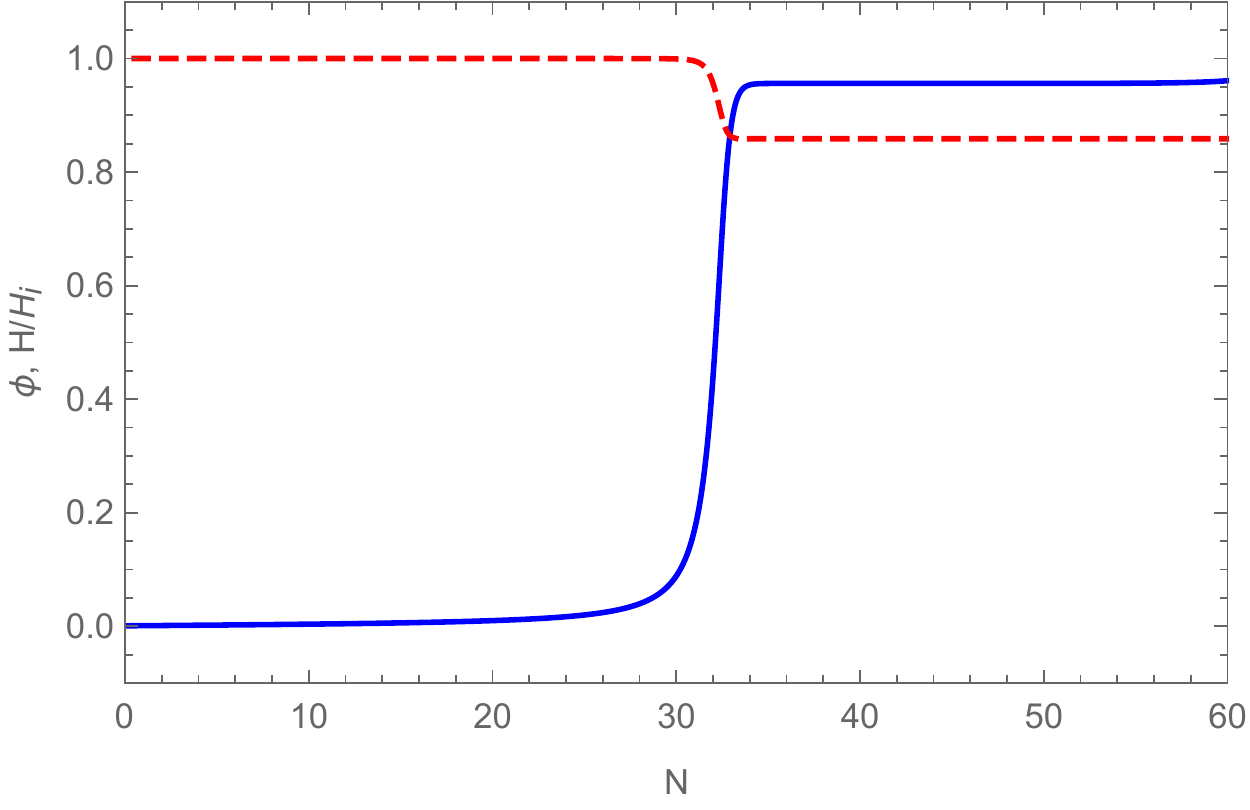}
\caption{Evolution of $H$ (dashed line) and $\phi$ (solid line) against e-folds $N$, with
{$H_i=7.65\times10^{-8}$}, {$\Phi_{0i}=8.42\times 10^{-4}$}, and {${\dot\Phi}_{0i}=1.43\times 10^{-11}$}.
Note that the inflation ends at $N\sim 60$.}
\label{figphiH}
\end{figure}

\begin{figure}[htp]
\centering
\includegraphics[width=0.6\textwidth]{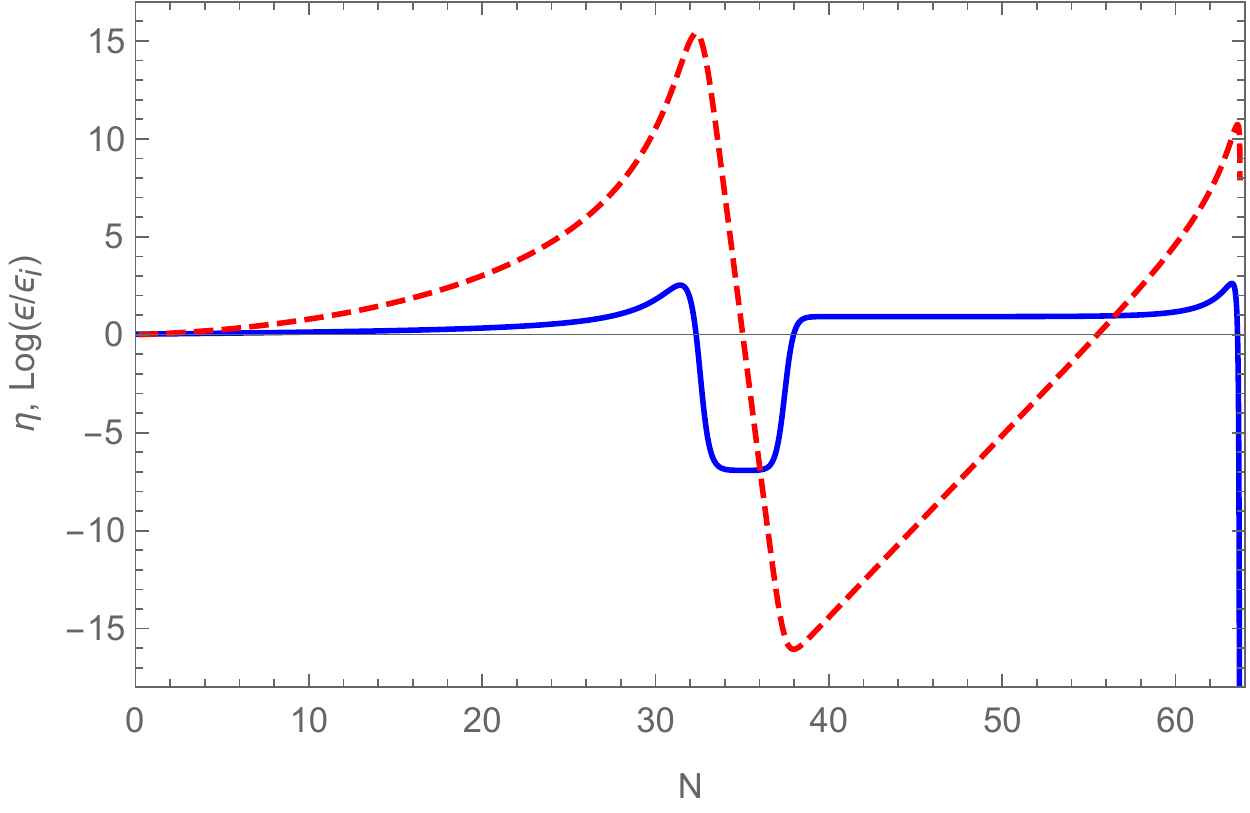}
\caption{Evolution of $\log (\epsilon / \epsilon_i)$ (dashed line) and $\eta$ (solid line) against e-folds $N$, where $\epsilon_i$ is the initial value of $\epsilon$.}
\label{figeps12}
\end{figure}

%Fig.5 zeta^2
\begin{figure}[htp]
\centering
\includegraphics[width=0.6\textwidth]{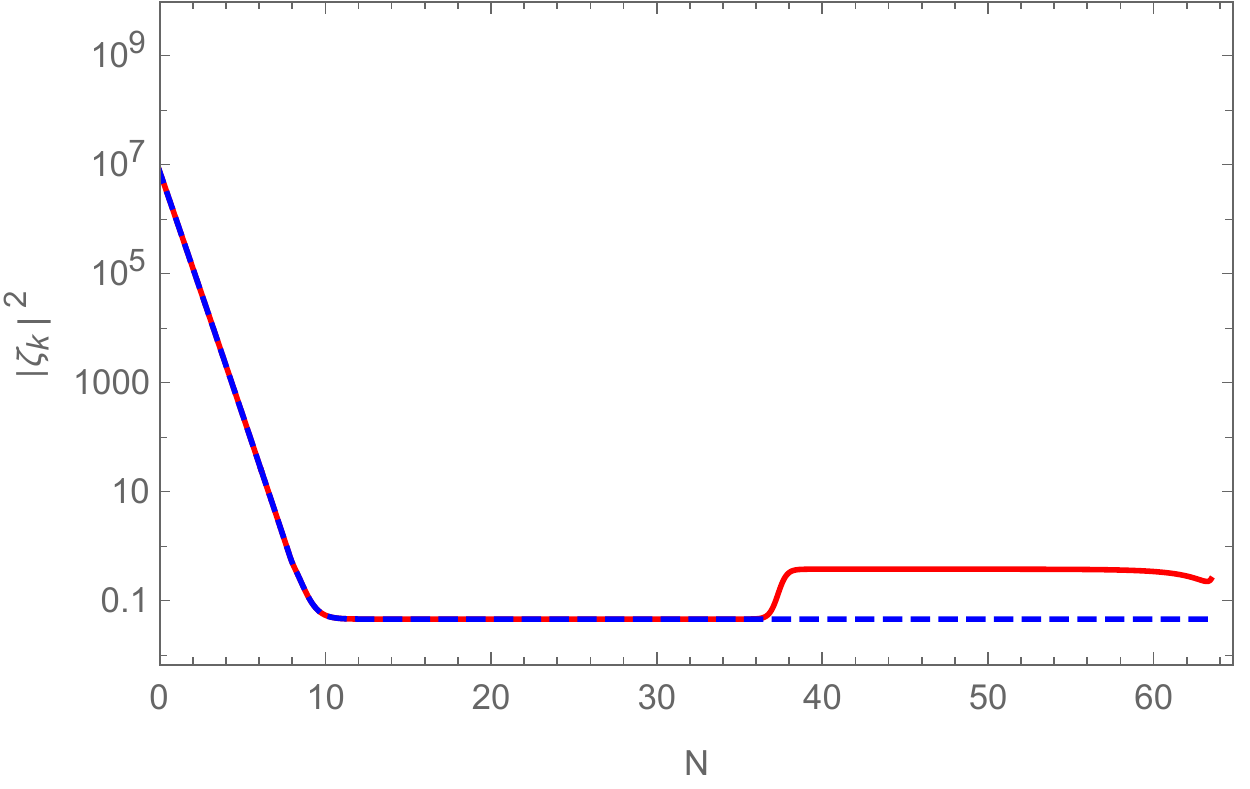}
\caption{{Evolution of $\left| \zeta_k \right|^2$ without back reaction (blue dashed line) and with back reaction (red solid line) against e-folds $N$, where $k/k_i=10^{4}$ and $k_i=H_i$.}}
\label{figzeta}
\end{figure}

%Fig.6 d zeta^2
\begin{figure}[htp]
\centering
\includegraphics[width=0.6\textwidth]{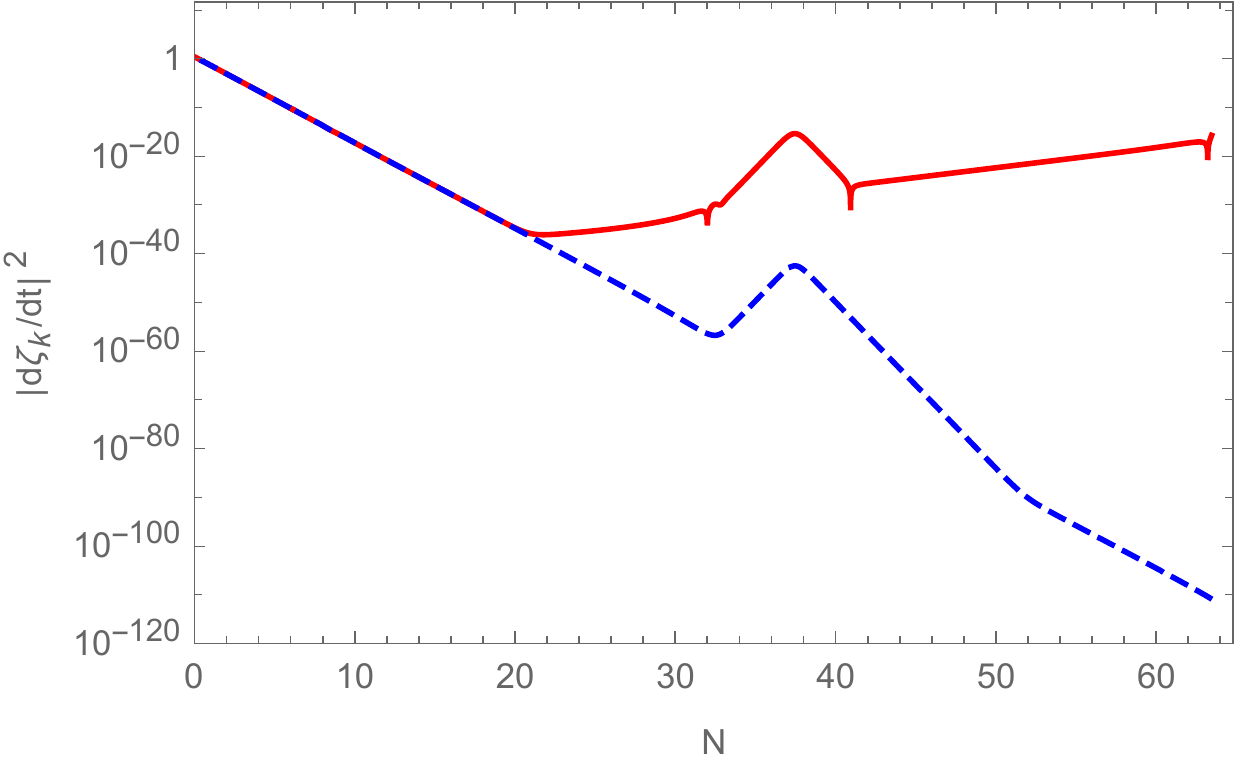}
\caption{{Evolution of $| \dot{\zeta}_k |^2$ without back reaction (blue dashed line) and with back reaction (red solid line) against e-folds $N$, where $k/k_i=10^{4}$ and $k_i=H_i$.}}
\label{figdzeta}
\end{figure}

\begin{figure}[htp]
\centering
\includegraphics[width=0.6\textwidth]{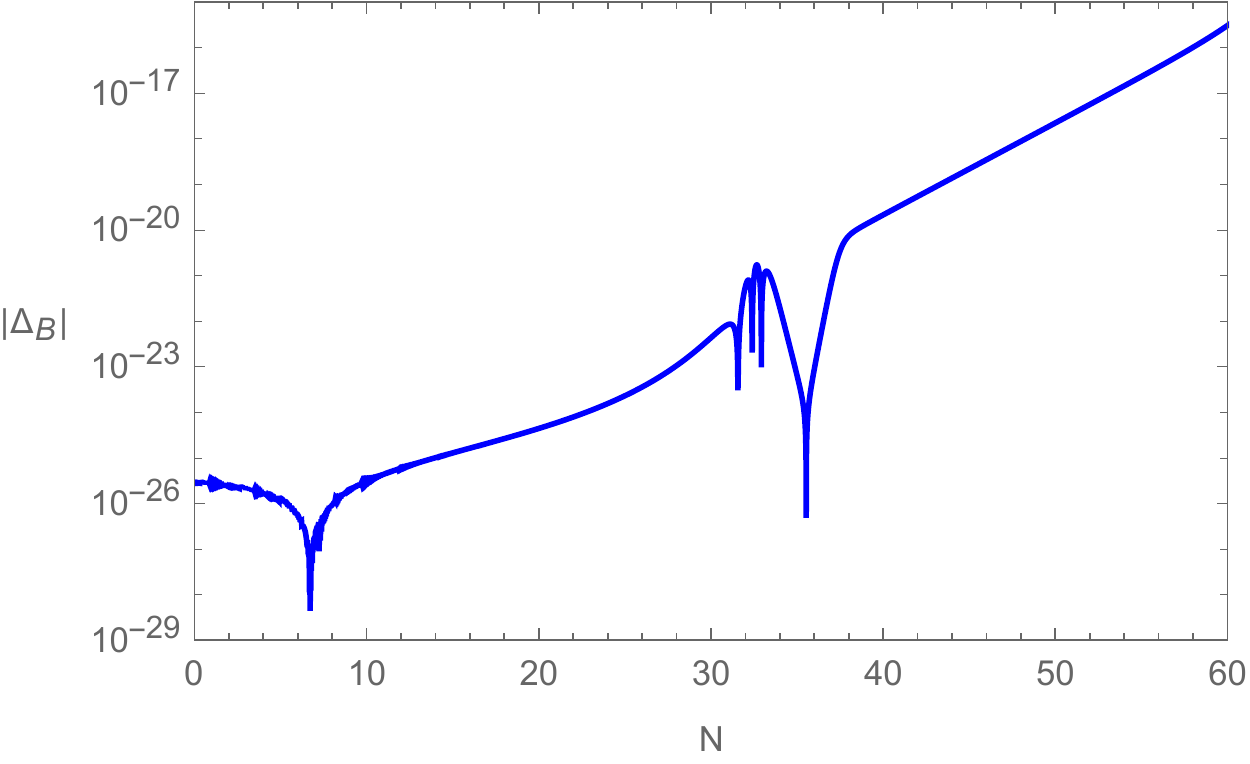}
\caption{{Evolution of $\Delta_B$ against e-folds $N$.}}
\label{figDeltaB}
\end{figure}
\begin{figure}[htp]
\centering
\includegraphics[width=0.6\textwidth]{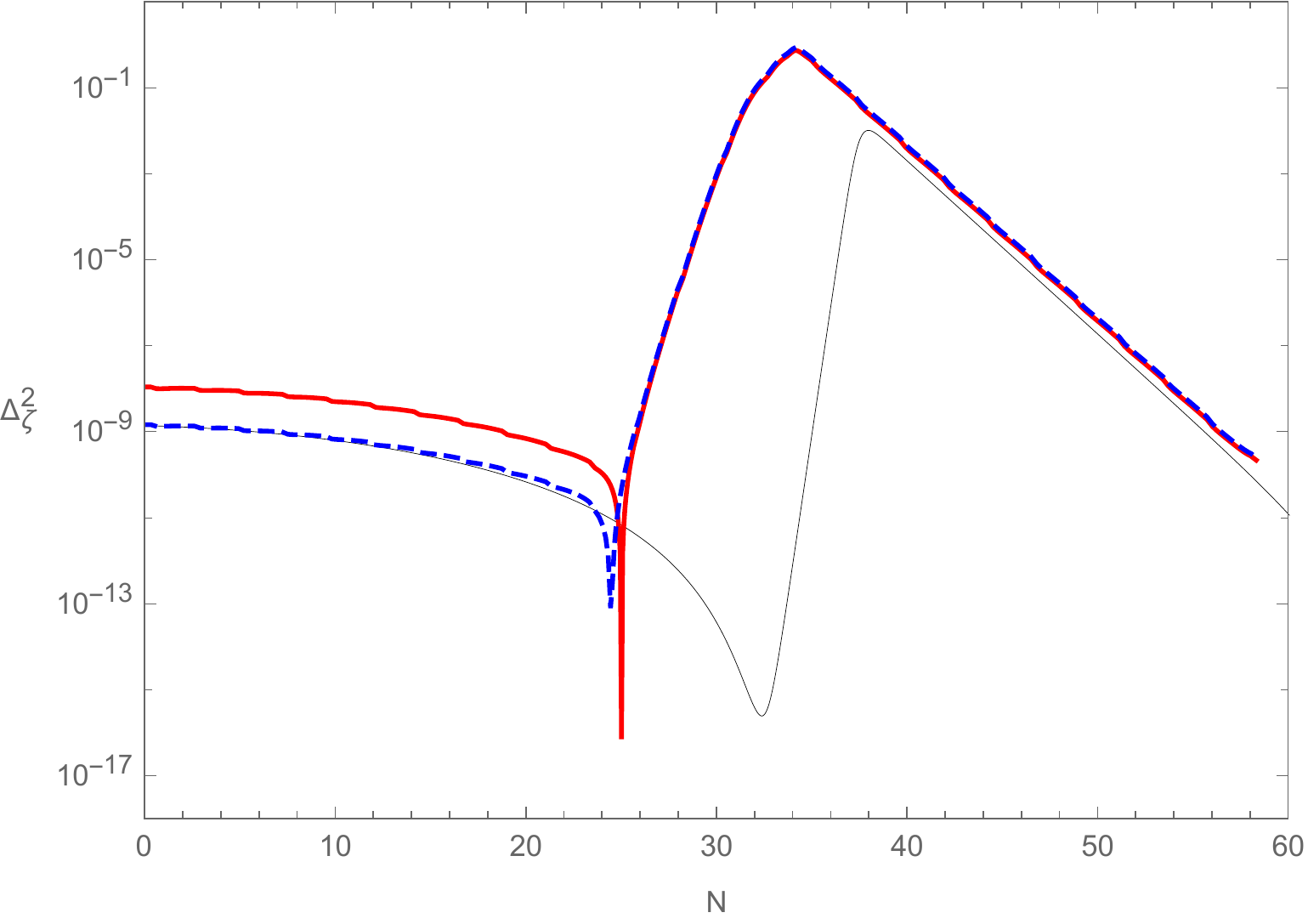}
\caption{{Power spectrum of the curvature perturbation $\Delta_{\zeta_k}$ without back reaction (blue dashed line), with back reaction (red solid line), and from the analytic result in Eq.~(\ref{deltaappro}) evaluated at the horizon-crossing time for each k-mode (black thin solid line) against e-folds $N$.}}
\label{figDeltaz}
\end{figure}

By taking account of the  quantum fluctuation effects, the solutions for the mean field $\Phi_0$,  the Hubble parameter $H$ and the modes of quantum field fluctuations $\zeta_k$
can be found by self-consistently solving the background field equations in (\ref{phi_loop}), and the Friedmann equation (\ref{frieman_q})  with the loop corrections,  together with the mode equations (\ref{MS2a})  that also include the contributions from the quantum field fluctuations themselves. The initial conditions of the mean field are set to be {$\Phi_{0 i} = 8.4 \times 10^{-4}$ and $ {\dot\Phi}_{0 i}= 1.43 \times 10^{-11}$} and those of the mode functions are  the  standard Bunch-Davies vacuum states.
The evolution of the mean field $\Phi_0$ and Hubble parameter $H$ is shown in  Fig.~\ref{figphiH} against
the number of e-folds after the beginning of inflation defined by $N=\int^t_0 H\left( t'\right) dt'$.
According to (\ref{Hubble_flow}), the Hubble flow parameters $\epsilon$ and $\eta$ are drawn in Fig.~\ref{figeps12}. It turns out that the mean field $\Phi_0$ starts at  $\Phi_{0i} \sim 10^{-4}$, and  slowly rolls down the potential hill in Fig.~\ref{figpotential}  to $\Phi_0 \sim 0.1$ in  {$N\sim 0-33$}
 where the Universe is in the slow-roll regime with small values of $\epsilon$ and $\eta$. Then, the  slope of the potential profile  drives the mean field quickly settling into another phase of the ultra-slow regime where $\Phi_0$ is sitting around the inflection point of the effective potential at $\Phi_0 \sim 0.9$ also seen in Fig.~\ref{figphiH}.
  There exists a sharp SR to USR transition at {$ N\sim 33$} in  Fig.~\ref{figeps12} where a relatively large change in $\epsilon$  results in the change in the  value of $\eta$ to $\eta \sim -7$ at {$N\sim 33$}, entering the new phase of the USR with the ignorable acceleration of the dynamics of the mean field. The USR lasts within {$N\sim 33 - 40$} when $\vert \eta \vert$ remains relatively large. Later, the value $\vert \eta \vert$ decreases and $\eta$ becomes positive. This in turn increases the value of $\epsilon$ and brings the phase back to the SR regime. Finally, the inflation phase has a exit at $N \sim 60$.

It is also found that although in the course of the Universe inflation and in  the models of interest,  the backreaction of $\langle \varphi^2 \rangle$
does not play a significant role in the evolution of the background fields. Instead, they indeed give visible effects in the mode functions $\zeta_k$ for relatively small $k$ modes,  which will affect the power spectrum of the curvature perturbations
for superhorizon modes.  The numerical results of $\left|\zeta_k \right|^2$ and  $|\dot{\zeta}_k |^2$ as a function of $N$ are displayed in Figs.~\ref{figzeta} and~\ref{figdzeta} respectively with a Fourier mode $k/k_i=10^4$ for illustration.
They are the relatively small $k$ modes with $k_i$ and $k$ corresponding to the modes crossing out of the horizon at the start of the inflation and at the early time  ${N\sim 8}$, respectively.
We find the large boost effect due to quantum fluctuations to the mode function $\zeta_k$, starting from  $N\sim 40$ onwards until the end of the inflation at $N\sim 60$.
This deviation can be ascribed to the nonzero $\Delta_B$ in the mode equations (\ref{MS2a})  where $\Delta_B=0$ when ignoring quantum fluctuations. As shown in Fig.~\ref{figDeltaB},
the value of $\vert \Delta_B \vert$ starts to grow from the onset of the inflation, drops its value at the SR-USR transition. It then hugely regrows after entering the USR phase with its effects to trigger the boost in the mode functions $\zeta_k$ until the end of the inflation while the Hubble parameter serving  as the cosmic friction effect is  not large enough to hinder the boost.
Given the equations for $\zeta_k$ in (\ref{MS2a}), evidently two competing factors, which are  the growth of $\Delta_B$ solely due to the quantum field fluctuations and the cosmic friction given by  the Hubble parameter $H$, will determine  whether or not the backreaction effects are visible in the model of interest.
In the first model above, the  boost effects occur for the modes  $k/k_i \lesssim 10^{11}$, which satisfy $k^2/a^2 \ll \vert \Delta_B\vert $ at $N=60$  in order to receive the significant effects from the nonzero $\Delta_B$ seen in the mode equations (\ref{MS2a}). What we learn is as follows. The need of the USR inflation is to generate the large enough fluctuations at relevant scales becoming the potential sources of the primordial black hole production. Nevertheless,  the created fluctuations will  backreact the dynamics of the mode functions if they are treated self-consistently as in our approach. This results in the sizable boost effects in particular on the small $k$ modes. As a consequence,  Fig.~\ref{figDeltaz} shows the power spectrum of the curvature perturbation $\Delta_{\zeta_k}$ with (red solid line) and without (blue dashed line) backreaction effects due to quantum fluctuations. This is one of the main results in our work.
As expected, the backreaction effects will induce noticeable deviation
 in small $k$ modes in particular when the Universe has the USR inflation.
 The range of such low-$k$ modes self-consistently depends on the value of $\Delta_B$ solely given by quantum fluctuations. In the second model we propose in this paper, the Hubble parameter can be sufficiently large so that the boost effects driven by the quantum fluctuations turn out to be insignificant.

\subsection{Model 2}
\begin{figure}[htp]
\centering
\includegraphics[width=0.6\textwidth]{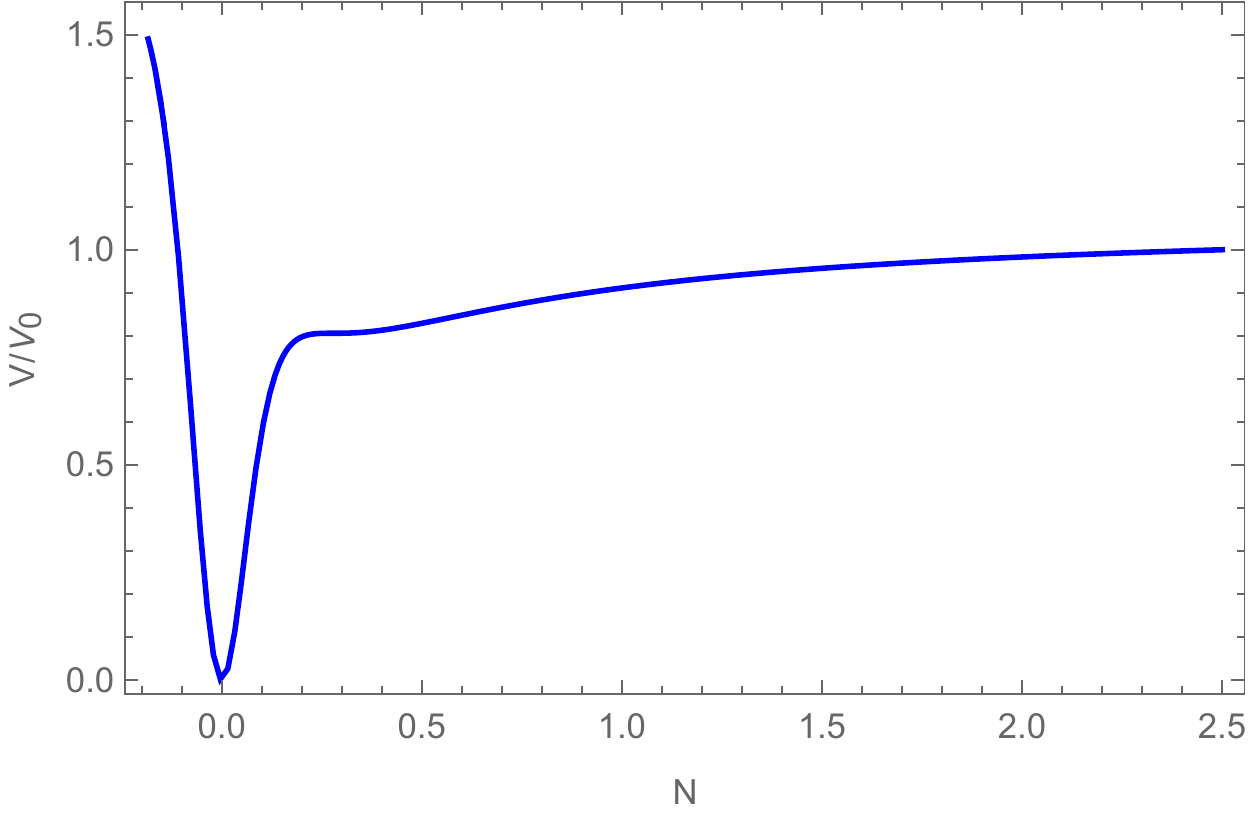}
\caption{The inflaton potential $V(\phi)$ in Eq.~(\ref{potential2}) for model 2 and $V_0=V(\Phi_{0i})$. All dynamical variables in this figure and in the following figures are rescaled by the reduced Planck mass, $M_p=2.435\times 10^{18}$ GeV. }
\label{figpotential_2}
\end{figure}

\begin{figure}[htp]
\centering
\includegraphics[width=0.6\textwidth]{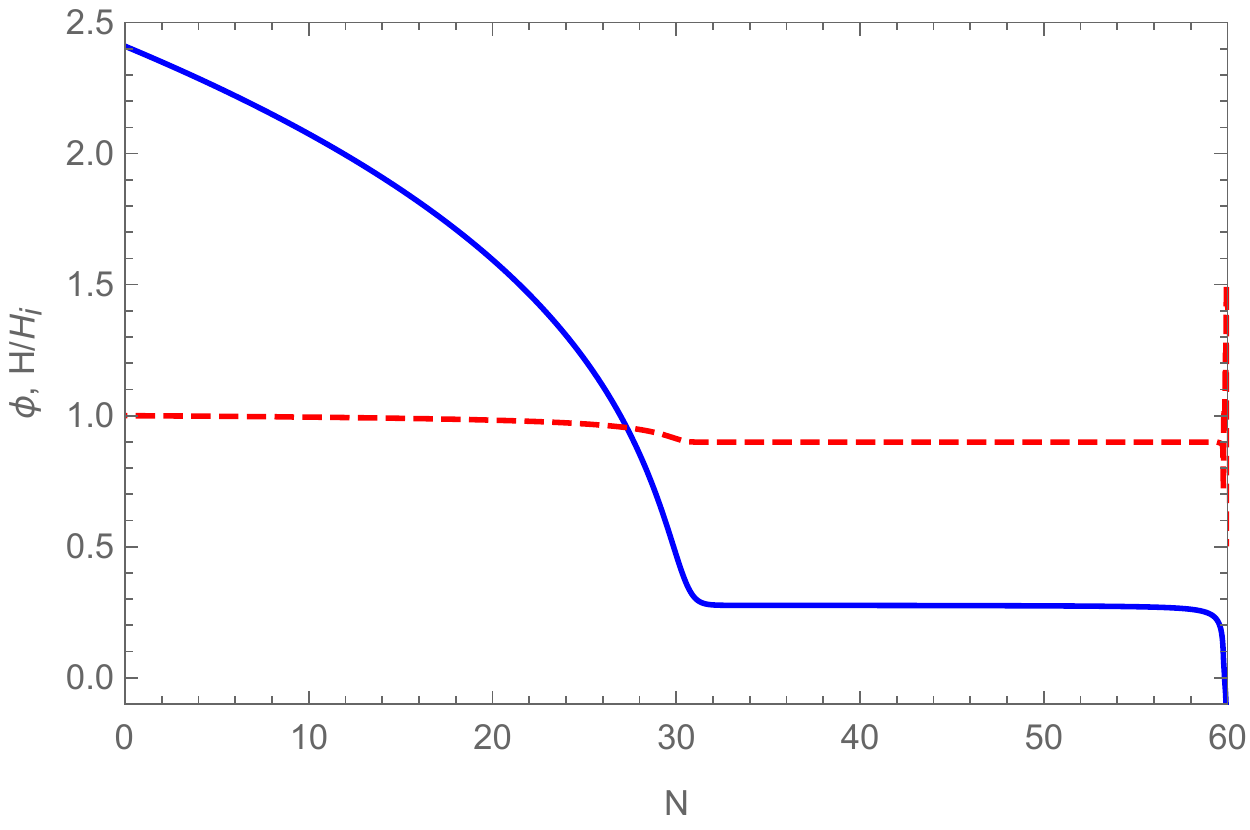}
\caption{{Evolution of $H$ (dashed line) and $\phi$ (solid line) against e-folds $N$, with
{$H_i=9.72\times 10^{-6}$,} $\Phi_{0i}=2.41$, and ${\dot\Phi}_{i}=-2.86\times10^{-7}$ for model 2.
Note that the inflation ends at $N\sim 60$.}}
\label{figphiH_2}
\end{figure}

\begin{figure}[htp]
\centering
\includegraphics[width=0.6\textwidth]{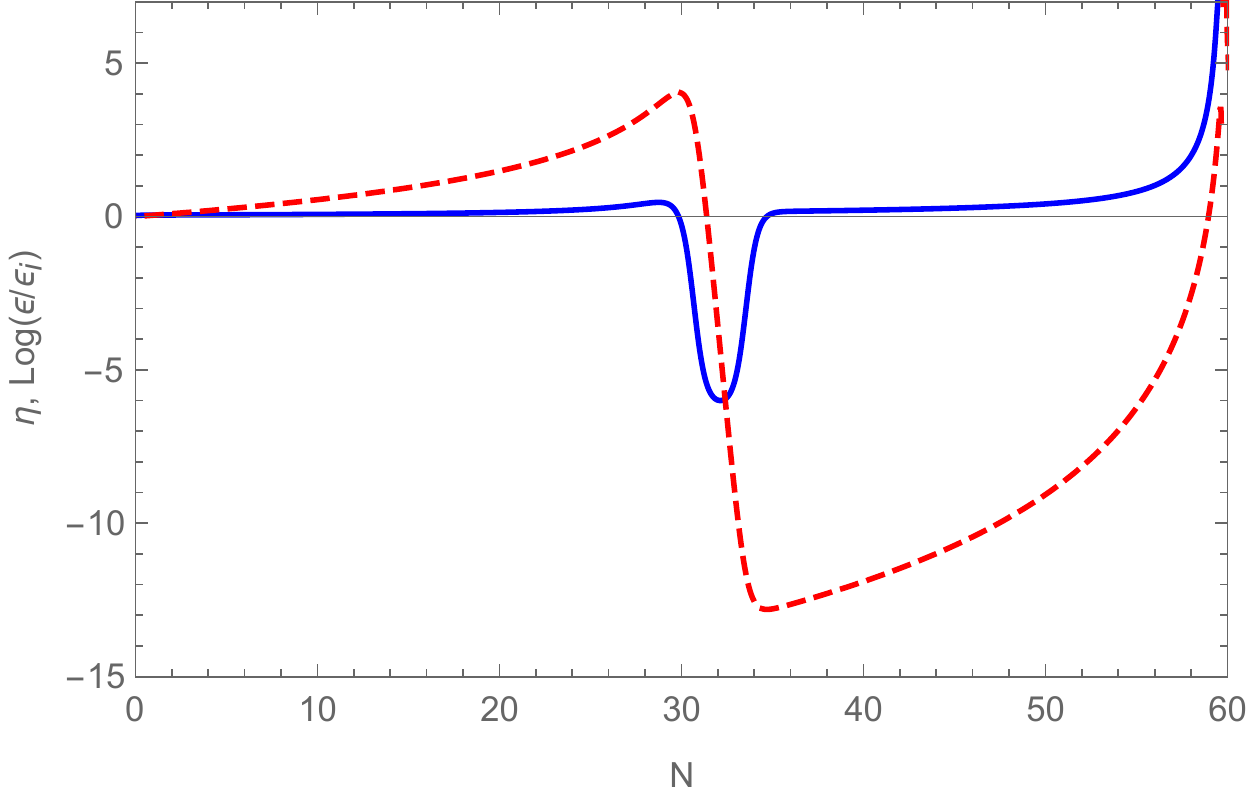}
\caption{Evolution of $\log (\epsilon / \epsilon_i)$ (dashed line) and $\eta$ (solid line) against e-folds $N$ for model 2, where $\epsilon_i$ is the initial value of $\epsilon$.}
\label{figeps12_2}
\end{figure}

\begin{figure}[htp]
\centering
\includegraphics[width=0.6\textwidth]{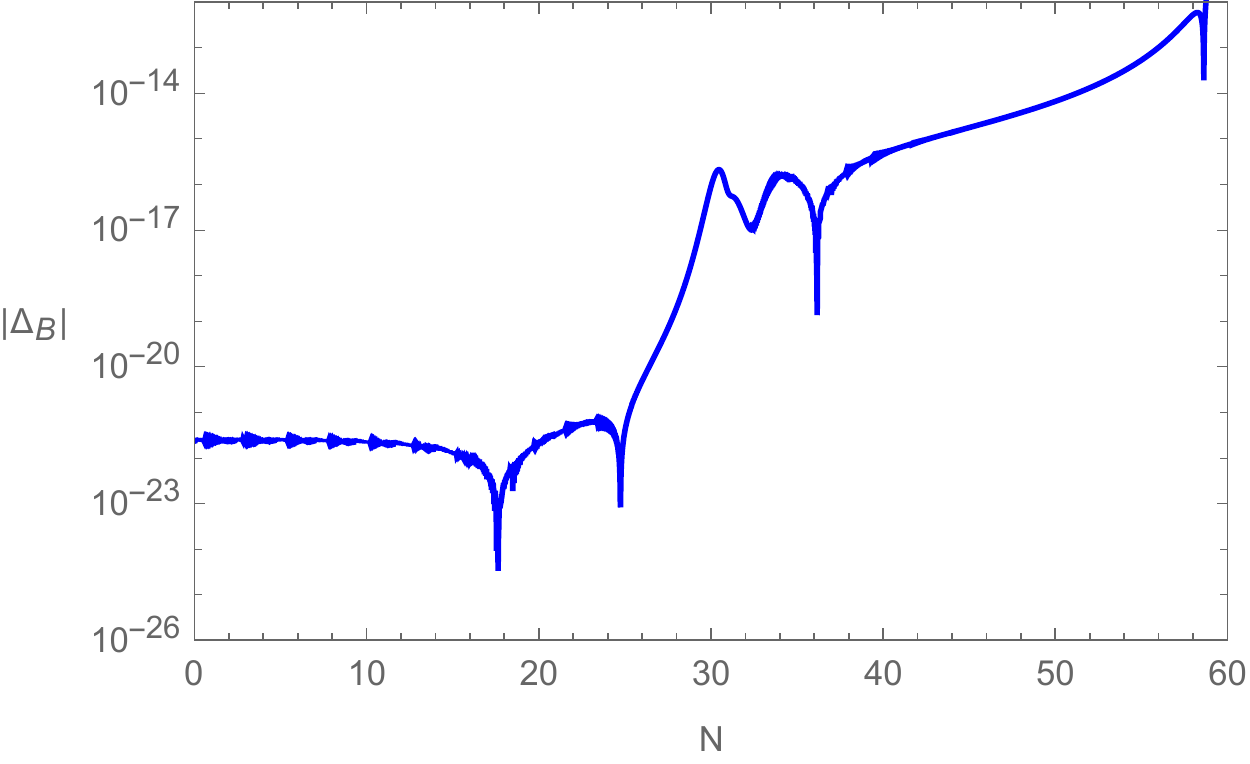}
\caption{{Evolution of $\Delta_B$ against e-folds $N$ for model 2.}}
\label{figDeltaB_2}
\end{figure}

\begin{figure}[htp]
\centering
\includegraphics[width=0.6\textwidth]{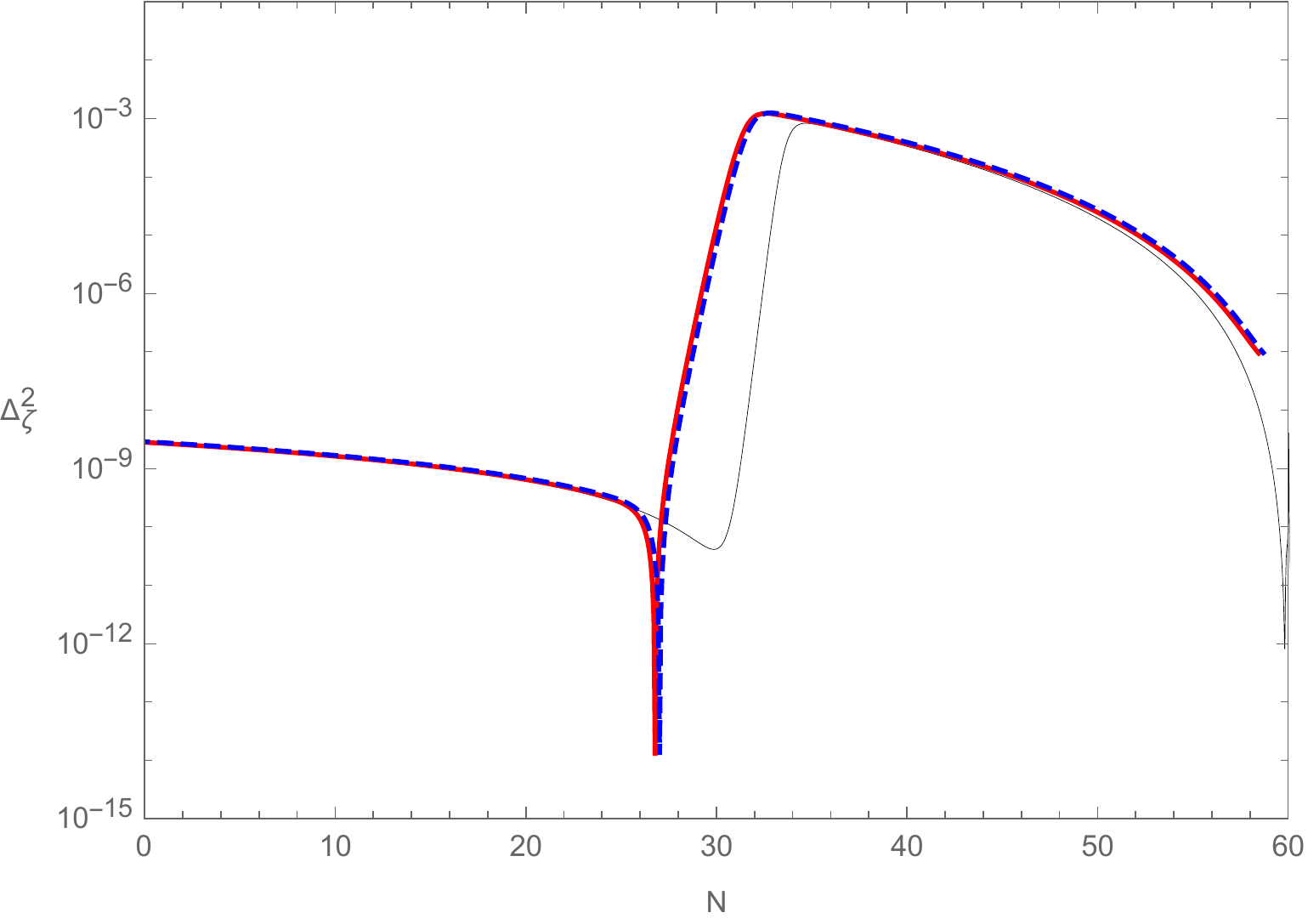}
\caption{{Power spectrum of the curvature perturbation $\Delta_{\zeta_k}$ without back reaction (blue dashed line), with back reaction (red solid line), and from the analytic result in Eq.~(\ref{deltaappro}) evaluated at the horizon-crossing time for each k-mode (black thin solid line) against e-folds $N$ for model 2. The dashed and solid curves overlap each other.}}
\label{figDeltaz_2}
\end{figure}

The second model we propose to study is parametrized as~\cite{EZQ}
\begin{equation}
V(\phi) = \frac{\lambda}{12} \phi^2 \nu^2 \frac{6-4 a \frac{\phi}{\nu} + 3 \frac{\phi^2}{\nu^2}}{\big[1+b \frac{\phi^2}{\nu^2} \big]^2} ,
\label{potential2}
\end{equation}
where
\begin{eqnarray}
{\lambda = 1.86 \times 10^{-6},\quad \nu=0.196, \quad a =0.7071, \quad b=1.5\,,}
\end{eqnarray}
and the corresponding potential profile is shown in Fig.~\ref{figpotential_2}.
To start from the SR inflation, the initial conditions for the mean field are chosen to be
{$\Phi_{0i}=2.41 $ and $\dot\Phi_{0i}=-2.86\times10^{-7}$} and with the parameters above in the model the Hubble parameter $H_i=9.71571\times 10^{-6}$. Again, without inclusion of the backreaction effects, the use of the chosen  parameters can successfully interpret the observations~\cite{planckinflation,bk18}. Applying the nonperturbative approach we develop can examine the quantum field fluctuation effects. The evolution of the background field $\Phi_0$ and the Hubble parameter are shown in Fig.~\ref{figphiH_2} although  the inclusion of $\langle \zeta^2 \rangle$  just gives the insignificant change in their evolution as in model 1. The evolution of the Hubble {flow} parameters is also depicted in Fig.~\ref{figeps12_2} where
the SR inflation starts from {$N= 0$} and ends at $N\sim 60$. As in model 1, there exists the SR-USR-SR transition and thus  $\Delta_B$ evolves similarly as shown in Fig.~\ref{figDeltaB_2}, which in particular largely regrows during the transition period, potentially giving the boost to $\zeta_k$.
However, in model 2, the Hubble parameter is  sufficiently large to prohibit the growth in $\zeta_k$ for all modes, resulting in the minute effects from the quantum field fluctuations to  $\Delta_{\zeta_k}$ shown in Fig.~\ref{figDeltaz_2}. In this work, we provide a self-consistent method to incorporate the quantum field fluctuations, which is a necessity to examine the backreaction effects from them.
 We will leave the study to find out in a given model the critical value of the Hubble parameter and its time-dependent behavior, below which the corrections from the quantum field fluctuations become of importance to our future work.  \\

%Fig.4 Delta_zeta^2

%Fig.2 phi and H

\begin{comment}
%Fig.5 zeta^2
\begin{figure}[htp]
\centering
\includegraphics[width=0.6\textwidth]{figzeta_2}
\caption{Evolution of $\left| \zeta_k \right|^2$ without back reaction (blue dashed line) and with back reaction (red solid line) against e-folds $N$ for model 2, where $k/k_i=10^{4}$ and $k_i=H_i$.}
\label{figzeta_2}
\end{figure}

%Fig.6 d zeta^2
\begin{figure}[htp]
\centering
\includegraphics[width=0.6\textwidth]{figdzeta_2}
\caption{{Evolution of $| \dot{\zeta}_k |^2$ without back reaction (blue dashed line) and with back reaction (red solid line) against e-folds $N$ for model 2, where $k/k_i=10^{4}$ and $k_i=H_i$.}
\label{figdzeta_2}
\end{figure}
\end{comment}

\begin{comment}
This deviation arise from the effect of the back reaction, $\Delta_B$.
During the  $\epsilon_2<0$ region, $\zeta_k$ modes will undergoing an anti-damping process and leads to an $\mathcal{O}(10^8)$ enhancement in $\dot{\zeta_k}$ for small $k$ modes where the wavenumber $k$ is smaller than the dip in the $\Delta_{\zeta_k}^2$ ($k/k_0 \sim 10^{11}$).
But without the back reaction, the amplified $\dot{\zeta}_k$ value is still too small so that the change of $\zeta_k$ is undistinguishable.
Due to the back reaction, $\dot{\zeta}_k$ become larger before this enhancement region, the deviation value that $\dot{\zeta}_k \Delta t \approx \dot{\zeta}_k \left( \Delta N / H \right)$ becomes comparable with $\zeta_k$, where $H \simeq 5\times 10^{-8}$, $\Delta N \simeq 1$, and $| \dot{\zeta}_k / \zeta_k | \simeq 6\times 10^{-8}$ at $N\sim 40$ for small $k$.
\end{comment}

\section{summary and looking ahead}\label{sec4}
In this work, we develop a nonperturbative method to   examine the  quantum fluctuation effects on the single-field inflationary models in a spatially flat FRW cosmological space-time.
 {We mainly follow the work of \cite{weican}
starting from the metric of the perturbed  spatially flat FRW cosmological space-time in the ADM form, and then  separate  the classical homogeneous background field $\Phi_0$ from the
quantum field fluctuations $\varphi$.   In a spatially flat gauge, the equation of motion for the background field  in the FRW metric as well as the modified Friedmann equation of the scale factor with quantum fluctuation corrections are derived.
 We adopt the Hartree factorization and treat the quantum field fluctuations $\langle \varphi^2 \rangle$ in a self-consistent manner in which the
 the equations of motion for mode functions of the quantum field $\varphi$ are also derived.
 In there the new variable $\Delta_B$ depending on not only the background field $\Phi_0$ and the Hubble parameter $H$ but also the quantum field fluctuations $\langle \varphi^2 \rangle$ is introduced.  In particular $\Delta_B$  vanishes according to the background field equation and the Friedmann equation as long as all quantum fluctuations are ignored. This reduces the mode equation to  the known Mukhanov-Sasaki equation.  Thus, $\Delta_{B} \neq 0$ is the indicator, giving the effects from the quantum fluctuations to the curvature perturbations, which  in a spatially flat gauge can be computed from the solutions of the mode functions.
We then consider the models in which
 the Universe undergoes  the SR-USR-SR inflation where with no inclusion of $\langle \varphi^2 \rangle$,
 the created curvature perturbations are consistent with the observations and also show a spike
 at relevant scales as the source for the primordial black hole production.
 Nevertheless,   although quantum fluctuations
do not play a significant role in the evolution of the background fields, they trigger the huge growth of
$\Delta_B$ in particular during the transition from USR back to the second SR inflation.
As a result, the growth of $\Delta_B$ in turn gives the boost effects to the curvature perturbations to the modes that leave the horizon in the early times of the  SR inflation.
Whether or not the boost effects can be visible  also depends on the cosmic friction term given by the Hubble parameter in the mode equations. Here we present two models to illustrate the competition between the boost and the friction force effects. In our future work, we plan to further study the critical value of the Hubble parameter and its time-dependent behavior, below which the corrections from the quantum field fluctuations become of importance in a given model of interest.
Thus, it will be also of great interest to construct a successful model to interpret the
observations and generate large enough spike for the PBH production where the accompanying large quantum fluctuations need to be taken into account.

\begin{acknowledgments}
This work was supported in part by the Ministry of Science and Technology (MOST) of Taiwan, R.O.C. under
grant numbers 109-2112-M-259-003 (DSL) and 109-2112-M-001-003 (KWN).
\end{acknowledgments}


\begin{references}
\bibitem{KT}
E. W. Kolb , M. S. Turner, The Early Universe Addison Wesley, Redwood City, C.A. 1990.
\bibitem{ABB}
B. P. Abbott {\it et al.} , Phys. Rev. Lett. {\bf 116}, 061102 (2016).

\bibitem{CAR}
B. J. Carr and S. W. Hawking, Mon. Not. Roy. Astron. Soc. {\bf 168}, 399 (1974).

\bibitem{MES}
P. Meszaros, Astron. Astrophys. {\bf 37}, 225 (1974).

\bibitem{CAR1}
 B. J. Carr, Astrophys. J. {\bf 201}, 1 (1975).

\bibitem{SAS}
M. Sasaki, T. Suyama, T. Tanaka, and S. Yokoyama, Class. Quant. Grav. {\bf 35}, 063001 (2018).

\bibitem{CHENG}
S.-L. Cheng, W.~Lee, and K.-W.~Ng,
Production of high stellar-mass primordial black holes in trapped inflation,
J. High Energy Phys. {\bf 02}, 008 (2017).

\bibitem{EZQcrit}
J. M. Ezquiaga, J. Garcia-Bellido, and E. R. Morales, Primordial Black Hole production in
Critical Higgs Inflation, Phys. Lett. B {\bf 776},  345  (2018).



\bibitem{KAN}
 K. Kannike, L. Marzola, M. Raidal, and H. Veerm{\"a}e, Single Field Double Inflation and
Primordial Black Holes, JCAP {\bf 1709}, 020 (2017).

\bibitem{BAL}
G. Ballesteros and M. Taoso, Phys. Rev. D {\bf 97}, 023501 (2018).

\bibitem{CLN}
S.-L. Cheng, W.~Lee, and K.-W.~Ng,
Primordial black holes and associated gravitational waves in axion monodromy inflation,
JCAP, {\bf 07}, 001 (2018).

\bibitem{CIC}
 M. Cicoli, V. A. Diaz, and F. G. Pedro, Primordial Black Holes from String Inflation, JCAP {\bf 06}, 034 (2018).


\bibitem{OZS}
 O. {\"O}zsoy, S. Parameswaran, G. Tasinato, and I. Zavala, Mechanisms for Primordial Black Hole
Production in String Theory, JCAP  {\bf 07},  005 (2018).


\bibitem{DAL}
 I. Dalianis, A. Kehagias, and G. Tringas, Primordial Black Holes from alpha-attractors, JCAP  {\bf 01}, 037 (2019).


\bibitem{KIN}
W. H. Kinney, Horizon crossing and inflation with large eta, Phys. Rev. D {\bf 72}, 023515 (2005).

\bibitem{MAR}
J. Martin, H. Motohashi, and T. Suyama, Ultra Slow-Roll Inflation and the non-Gaussianity Consistency Relation, Phys. Rev. D {\bf 87}, 023514 (2013).

\bibitem{CHE}
S.-L. Cheng, W. Lee, and K.-W. Ng, Superhorizon curvature perturbation in ultraslow-roll inflation,
Phys. Rev. D {\bf 99}, 063524 (2019).

\bibitem{BYR}
C. T. Byrnes, P. S. Cole, and S. P. Patil, Steepest growth of the power spectrum and primordial black holes, JCAP {\bf 06}, 028 (2019).

\bibitem{STA}
A. A. Starobinsky, Stochastic De Sitter (inflationary) stage in the early Universe, Lect. Notes Phys. {\bf 246}, 107 (1986).


\bibitem{PAT}
 C. Pattison, V. Vennin, H. Assadullahi, and D. Wands, Quantum diffusion during inflation and primordial black holes, JCAP {\bf 10}, 046 (2017); Stochastic inflation beyond slow roll, JCAP {\bf 07}, 031 (2019).


\bibitem{BIA}
M. Biagetti, G. Franciolini, A. Kehagias, and A. Riotto, Primordial Black Holes from Inflation and Quantum Diffusion, JCAP {\bf 07},  032 (2018).

\bibitem{EZQ}
J.  M. Ezquiaga and J. Garcia-Bellido, Quantum diffusion beyond slow-roll: implications for primordial black-hole production, JCAP {\bf 08}, 018 (2018).

\bibitem{FIR}
H. Firouzjahi, A. Nassiri-Rad, M.  Noorbala, Stochastic Ultra Slow Roll Inflation,
JCAP {\bf 01}, 040 (2019).



\bibitem{boyan1}
D. Boyanovsky, H. J. de Vega, and N. G. Sanchez,
Nucl. Phys. B {\bf 747}, 25 (2006).

\bibitem{fordbunch} A. Vilenkin and L. H. Ford, Phys. Rev. D
{\bf 26}, 1231 (1982); T. S. Bunch and P. C. W. Davies, Proc. R.
Soc. London A  {\bf 360}, 117 (1978).


\bibitem{weican}
W.-C. Syu, D.-S. Lee, and K.-W. Ng, Quantum loop effects to the power spectrum of primordial perturbations during ultra slow-roll inflation,
Phys. Rev. D {\bf 101}, 025013 (2020).


\bibitem{clu}
 T. Clunan and D. Seery, Relics of spatial curvature in the primordial non-gaussianity, JCAP {\bf 1001}, 032 (2010).
\bibitem{kaz}
 K. Sugimura and E. Komatsu, Bispectrum from open inflation
, JCAP {\bf 11}, 065 (2013).

\bibitem{boyan3}  D. Boyanovsky,  H. J. de Vega,  R. Holman, D.-S.  Lee
and   A. Singh, Phys.  Rev. D {\bf 51}, 4419 (1995);   D. Boyanovsky, M. D'Attanasio,  H. J. de Vega,  R. Holman and
D.-S.  Lee,  Phys. Rev. D {\bf 52}, 6805
  (1995).

\bibitem{msvar}
M. Sasaki, Prog. Theor. Phys. {\bf 76}, 1036 (1986);
V. F. Mukhanov, Zh. Eksp. Teor. Fiz. {\bf 94N7}, 1
(1988) [Sov. Phys. JETP {\bf 67}, 1297 (1988)].

\bibitem{planckinflation}
Planck Collaboration, Y. Akrami {\it et al.},Planck 2018 results. X. Constraints on inflation,  A \& A {\bf 641}, A10 (2020).

\bibitem{bk18}
Keck Array and BICEP2 Collaborations: P. A. R. Ade {\it et al.}, BICEP2 / Keck Array x: Constraints on Primordial Gravitational Waves using Planck, WMAP, and New BICEP2/Keck Observations through the 2015 Season, Phys. Rev. Lett. {\bf 121}, 221301 (2018).























\end{references}
\end{document}